\documentclass[a4paper,12pt]{article}

\usepackage{jcappub} 
\usepackage{hyperref}
\hypersetup{
    bookmarks=true,         
    unicode=false,          
    pdftoolbar=true,        
    pdfmenubar=true,        
    pdffitwindow=false,     
    pdfstartview={FitH},    
    pdftitle={My title},    
    pdfauthor={Author},     
    pdfsubject={Subject},   
    pdfcreator={Creator},   
    pdfproducer={Producer}, 
    pdfkeywords={keyword1} {key2} {key3}, 
    pdfnewwindow=true,      
    colorlinks=true,       
    linkcolor=red,          
    citecolor=cyan,        
    filecolor=magenta,      
    urlcolor=blue,           
    linktocpage=true
}
\usepackage{subfigure}
\usepackage{pstricks}
\usepackage{color}
\usepackage{amsfonts}
\usepackage{mathrsfs}
\usepackage{epsfig}
\usepackage{amsmath,amssymb,amsthm,graphicx,latexsym}
\usepackage{ulem}

\newcommand{\be}{\begin{equation}}
\newcommand{\ee}{\end{equation}}
\newcommand{\bea}{\begin{eqnarray}}
\newcommand{\eea}{\end{eqnarray}}
\newcommand{\bml}{\begin{subequations}}
\newcommand{\eml}{\end{subequations}}
\newcommand{\bfig}{\begin{figure}}
\newcommand{\efig}{\end{figure}}

\newcommand{\slashed}{\hspace{-1.1ex}/}

\title{
 Reconstructing  inflationary potential from BICEP2 and running of tensor modes}

\author[a]{Sayantan Choudhury}
\author[b]{ Anupam Mazumdar}

\affiliation[a]{Physics and Applied Mathematics Unit, Indian Statistical Institute, 203 B.T. Road, Kolkata 700 108, INDIA}
\affiliation[b]{Consortium for Fundamental Physics, Physics Department, Lancaster University, LA1 4YB, UK}

\abstract{ In this paper we will analyse the constraints on a sub-Planckian excursion of a single inflaton field, 
which would yield a large tensor to scalar ratio, while explaining the temperature anisotropy of the 
cosmic microwave background (CMB) radiation. In particular, our attempt will be to reconstruct the inflationary potential by 
constraining, $V(\phi_0),~V^{\prime}(\phi_0),~V^{\prime\prime}(\phi_0),~V^{\prime\prime\prime}(\phi_0)$ and 
$V^{\prime\prime\prime\prime}(\phi_0)$, in the vicinity of the field, $\phi_0\ll M_p$, and the field displacement, $\Delta \phi \ll M_p$, where 
$M_p$ is the reduced Planck mass. We will provide, for the first time, a set of new {\it consistency} relationships for sub-Planckian excursion
of the inflaton field, which would help us to differentiate sub-versus-super Planckian models of inflation. 
For a generic single field inflationary potential, we will be able to put a stringent bound on
the potential energy density: $2.07\times10^{16}~{\rm GeV}\leq\sqrt[4]{V_{\star}}\leq 2.40\times 10^{16}~{\rm GeV}$, where inflation
can occur on the flat potential within, 
$0.066 \leq\frac{\left |\Delta\phi\right|}{M_p}\,\leq 0.092$, for the following observational constraints: (Planck+WMAP-9+high L+BICEP2). We
then provide a prediction for the spectral tilt ($n_{T}$), running ($\alpha_{T}$) and running of running ($\kappa_{T}$) 
of the tensor modes within the window, $-0.019<n_{T}<-0.033$,  $-2.97\times 10^{-4}<\alpha_{T}(=dn_{T}/d\ln k)<2.86\times 10^{-5}$,
and $-0.11\times 10^{-4}<\kappa_{T}(=d^{2}n_{T}/d\ln k^{2})<-3.58\times 10^{-4}$, in a model independent way.
We also provide a simple example of an {\it inflection-point} model of inflation and reconstruct the potential in a model independent way to match the
current observations. 

}

\begin{document} 
\maketitle
\flushbottom

\section{Introduction}

The primordial inflation~\cite{Guth:1980zm,Linde,Albrecht} has two {\it key} predictions - creating the scalar density perturbations and the tensor perturbations during the accelerated phase of expansion~\cite{Mukhanov:1981xt}, for a review, see~\cite{Mukhanov:1990me}. One of the predictions, namely the temperature anisotropy due to the 
scalar density fluctuations has now been tested very accurately by the observations from the temperature anisotropy in the cosmic microwave background (CMB) radiation~\cite{Hinshaw:2012aka,Planck-1,Planck-infl}. 
The detection of  tensor modes has now recently been confirmed by the ground based BICEP experiment~\cite{Ade:2014xna}, which has detected
for the first time a non-zero value of the tensor-to-scalar ratio at $7\sigma$ C.L. The value obtained by the BICEP team in conjunction with
(Planck+WAMP-9+high L+BICEP2) put a bound on the primordial gravitational waves, via tensor-to-scalar ratio, within a window, 
$0.15 \leq r(k_\star)\equiv P_T(k_\star)/P_S(k_\star)\leq  0.27$, at the pivot scale, $k_\star = 0.002 {\rm Mpc^{-1}}$~\cite{Ade:2014xna}, where
$P_T$ and $P_S$ denote the power spectrum for the tensor and scalar modes, respectively. Note that large $r(k_\star)$ is possible only if the 
initial conditions for gravitational waves is quantum Bunch-Davis vacuum~\cite{BD}, for a classical initial condition the amplitude of the gravitational 
waves would be very tiny and undetectable~\cite{Ashoorioon:2012kh}, therefore the first observable proof of quantum gravity.

In this paper, our aim will be to illustrate that it is possible to explain the current data sets (Planck+WAMP-9+high L+BICEP2)
within a sub-Planckian VEV model of inflation, where:
\be
\phi_0 \ll M_p\,,~~~~~~|\Delta\phi |\approx |\phi_\star-\phi_e| \ll M_p,
\ee
where $\phi_\star \geq \phi_0\geq \phi_e$ represents the field VEV, and $\Delta\phi$ denotes the range of the field values around which all 
the relevant inflation occurs, $\phi_\star$ corresponds to the pivot scale and $\phi_e$ denotes the end of inflation, and 
$M_p=2.4\times 10^{18}$~GeV. Naturally, the potential has to be flat enough within $\Delta \phi$ to support slow roll inflation.

The above requirements are important if the origin of the inflaton has to be embedded within a particle theory, where inflaton is part of
a {\it visible sector} gauge group, i.e. Standard Model gauge group, instead of an arbitrary gauge singlet, for a review, see~\cite{Mazumdar:2010sa}. 
If the inflaton is {\it gauged} under some gauge group,
as in the case of a minimal supersymmetric Standard Model (MSSM), Ref.~\cite{Allahverdi:2006iq}, then the inflaton VEV must be 
bounded by $M_p$, in order to keep the sanctity of an effective field theory description~\footnote{ An arbitrary moduli or a gauge singlet inflaton can 
take large VEVs ( super-Planckian ) as in the case of a chaotic inflation~\cite{Linde}. Although, in the case of {\it assisted inflation}~\cite{assisted}, see {\it chaotic assisted inflation}~\cite{kanti},  the individual VEVs of the inflatons are sub-Planckian.}.

Our treatment will be very generic, with a potential given by:
\begin{eqnarray}\label{rt10a}
V(\phi)&=&V(\phi_0)+V^{\prime}(\phi_0)(\phi-\phi_{0})+\frac{V^{\prime\prime}(\phi_0)}{2}(\phi-\phi_{0})^{2}+\frac{V^{\prime\prime\prime}(\phi_0)}{6}(\phi-\phi_{0})^{3}\,\nonumber\\
&&~~~~~~~~~~~~~~~~~~~~~~~~~~~~~~~~~~~~~~~~~~~~~~~~+\frac{V^{\prime\prime\prime\prime}(\phi_0)}{24}(\phi-\phi_{0})^{4}+\cdots\,,
\end{eqnarray}
where $V(\phi_0)\ll M_p^4$ denotes the height of the potential, 
and the coefficients: $V^{\prime}(\phi_0) \leq M_p^3,~V^{\prime\prime}(\phi_0)\leq M_p^2,~V^{\prime\prime\prime}(\phi_0)\leq M_p,~V^{\prime\prime\prime\prime}(\phi_0)\leq {\cal O}(1)$, determine the shape of the potential in terms of the model parameters. The {\it prime} denotes the derivative w.r.t. $\phi$~\footnote{ In particular, some specific choices of the potential would be a {\it saddle point}, when  $V^{\prime}(\phi_0)=0=V^{\prime\prime}(\phi_0)$,  an  {\it inflection point}, when $V^{\prime\prime}(\phi_0)=0$.}.

Previous studies regarding obtaining large $r(k_\star)$ within sub-Planckian VEV models of inflation have been studied in 
Refs.~\cite{BenDayan:2009kv,Shafi,Hotchkiss:2011gz,Choudhury:2013iaa}. In Ref.~\cite{BenDayan:2009kv},
the authors could match the amplitude of the power spectrum, $P_S$, at the pivot point, but not at the entire range of $\Delta\phi$ for the observable window of $\Delta N$, where
$N$ is the number of e-foldings of inflation. In Ref.~\cite{Hotchkiss:2011gz}, the authors have looked into higher order slow roll corrections by expanding the  potential 
around $\phi_0$. They pointed out that large $r\sim 0.05$ could be obtained in an {\it inflection-point} model of inflation where the slow roll parameter, $\epsilon_V$, changes 
non-monotonically ( for a definition of $\epsilon_V$, see Eq.~(\ref{ra1}) ). The $\epsilon_V$ parameter first increases within the observational window of $\Delta N$ and then decreases before increasing to exit the slow roll inflation by violating the slow roll condition, i.e. $\epsilon_V\approx 1$.
The value of $r$ was still small in order to accommodate the WMAP data, which had probed roughly $ \Delta N\approx 8$ as compared to the Planck, which has now probed $\Delta N\approx 17$ e-foldings of inflation.

Following Ref.~\cite{Hotchkiss:2011gz}, a generic bound on tensor to scalar ratio was presented in Ref.~\cite{Choudhury:2013iaa} for an {\it inflection-point} model only,
when $V^{\prime\prime}(\phi_0)=0$. 

In this paper we will consider the full potential of Eq.~(\ref{rt10a}), and our goal will be to determine the values 
of $V(\phi_\star),~V^{\prime}(\phi_\star),~V^{\prime\prime}(\phi_\star),~V^{\prime\prime\prime}(\phi_\star)$ and $V^{\prime\prime\prime\prime}(\phi_\star)$ from the current 
(Planck+WAMP-9+high L+BICEP2) data. In this respect we will be reconstructing the inflationary potential around $\phi_0$ and the pivot scale, $\phi_\star=\phi(k_\star)$, 
where $k_\star = 0.002 {\rm Mpc^{-1}}$. We will also provide for he first time the consistency relations for a sub-Planckian excursion of the inflaton field. This will provide us
the with an observational discriminator which could falsify sub-Planckian models of inflation in future.

Within Planck's observable region of  $\Delta {N}\approx 17$ e-foldings,  
we will be able to constrain the power spectrum: $P_S$, spectral tilt: $n_S$,
running of the spectral tilt: $\alpha_S$, and running of running of the spectral tilt: $\kappa_S$, 
for (Planck+WMAP-9+high~L+BICEP2) data sets:
 \begin{eqnarray}\label{obscons}
 0.15 &\leq & r(k_\star)\leq  0.27\\
 \ln(10^{10}P_{S})&=&3.089^{+0.024}_{-0.027}~~ ({\rm within}~ 2\sigma ~C.L.),\\
 n_{S}&=&0.9600 \pm 0.0071~~ ({\rm within}~ 3\sigma ~C.L.),\\
 \alpha_{S}&=&dn_{S}/d\ln k=-0.022\pm 0.010~~({\rm within}~1.5\sigma~C.L.),\\
 \kappa_{S}&=&d^{2}n_{S}/d\ln k^{2}=0.020^{+0.016}_{-0.015}~~({\rm within}~1.5\sigma~C.L.)\,.
 \end{eqnarray}

We will briefly recap the main equations for the tensor to scalar ratio in the most general case by taking into account of the higher order slow-roll conditions.
We will then derive the {\it most general} bound on $r(k_\star)$ for a sub-Planckian VEV inflation, and the corresponding values of $H_\star$ and $V(\phi_\star)$.
 We will then reconstruct the shape of the potential in Section 4 by providing the constraints on $V^{\prime}(\phi_0),~V^{\prime\prime}(\phi_0),~V^{\prime\prime\prime}(\phi_0)$
 and $V^{\prime\prime\prime\prime}(\phi_0)$. In 
Section 5, we will discuss the consistency relationships, and in section 6, we will consider a specific case of inflection point inflation for the purpose of illustration. 
We will provide our key equations in an Appendix.


\section{Brief discussion on tensor to scalar ratio} 

The tensor to scalar ratio can be  defined by taking into account of
the higher order corrections, see~\cite{Easther:2006tv,Choudhury:2013iaa,Choudhury:2013jya}:
\be\label{para 21e} 
r\approx16\epsilon_{V}\frac{\left[1-({\cal C}_{E}+1)\epsilon_{V}\right]^{2}}{\left[1-(3{\cal C}_{E}+1)\epsilon_{V}
+{\cal C}_{E}\eta_{V}\right]^{2}}\,
\ee
where ${\cal C}_{E}=4(\ln 2+\gamma_{E})-5$ with $\gamma_{E}=0.5772$ is the {\it Euler-Mascheroni constant}, and 
slow-roll parameters  $(\epsilon_{V},~\eta_{V})$ are given by in terms of the inflationary potential $V(\phi)$, which can be expressed as:
\begin{eqnarray}\label{ra1}
    \epsilon_{V}=\frac{M^{2}_{p}}{2}\left(\frac{V^{\prime}}{V}\right)^{2}\,,~~~~~~~
    \label{ra2} \eta_{V}={M^{2}_{P}}\left(\frac{V^{\prime\prime}}{V}\right)\,.
   \end{eqnarray}
We would also require two other slow-roll parameters, $(\xi^{2}_{V},\sigma^{3}_{V})$, in our analysis, which are given by:
\begin{eqnarray}\label{ja1}
    \xi^{2}_{V}=M^{4}_{p}\left(\frac{V^{\prime}V^{\prime\prime\prime}}{V^{2}}\right)\,,~~~~~~~~
    \label{ja2} \sigma^{3}_{V}=M^{6}_{p}\left(\frac{V^{\prime 2}V^{\prime\prime\prime\prime}}{V^{3}}\right)\,.
   \end{eqnarray}
Note that we have neglected the contributions from the higher order slow-roll terms, as they are sub-dominant at the leading order.
With the help of 
\be\label{con1}
 \frac{d}{d\ln k}=-M_p\frac{\sqrt{2\epsilon_{H}}}{1-\epsilon_{H}}\frac{d}{d\phi}\,\approx -M_p\frac{\sqrt{2\epsilon_{V}}}{1-\epsilon_{V}}\frac{d}{d\phi}\,,
\ee
we can derive a simple expression for the tensor-to-scalar ratio, $r$, as:~\footnote{ We have derived some of the key expressions in an Appendix, see for instance, Eq.~(\ref{para 21e}), which we would require to derive the above expression, Eq.~(\ref{con2}). }
\be\label{con2}
 r=\frac{8}{M^{2}_{p}}\frac{(1-\epsilon_{V})^{2}\left[1-({\cal C}_{E}+1)\epsilon_{V}\right]^{2}}{\left[1-(3{\cal C}_{E}+1)\epsilon_{V}
+{\cal C}_{E}\eta_{V}\right]^{2}}\left(\frac{d\phi}{d{\ln k}}\right)^{2}. 
 \ee
We can now derive a bound on $r(k)$  in terms of the momentum scale:
\be\begin{array}{llll}\label{con4}
    \displaystyle \int^{{ k}_{\star}}_{{k}_{e}}\frac{dk}{k}\sqrt{\frac{r({k})}{8}} \\ \displaystyle =
\frac{1}{M_p}\int^{{\phi}_{\star}}_{{\phi}_{e}}d {\phi}\frac{(1-\epsilon_{V})\left[1-({\cal C}_{E}+1)\epsilon_{V}\right]}{\left[1-(3{\cal C}_{E}+1)\epsilon_{V}
+{\cal C}_{E}\eta_{V}\right]},\\
 \displaystyle \approx \frac{1}{M_p}\int^{{\phi}_{\star}}_{{\phi}_{e}}d {\phi}(1-\epsilon_{V})\left[1+{\cal C}_{E}(2\epsilon_{V}-\eta_{V})+....\right],\\
\displaystyle \approx \frac{\Delta\phi}{M_p} \left\{ 1+\frac{1}{\Delta\phi}\left[(2{\cal C}_{E}-1)\int^{{\phi}_{\star}}_{{\phi}_{e}}d {\phi}~\epsilon_{V}
-{\cal C}_{E}\int^{{\phi}_{\star}}_{{\phi}_{e}}d {\phi}~\eta_{V}\right]+....\right\}\,,
   \end{array}\ee
where note that $\Delta\phi \approx \phi_{\star}-\phi_{e}>0$ is positive in Eq.~(\ref{con4}), and $\phi_e$ denotes the inflaton VEV at the end of inflation,
and $\phi_{\star}$ denote the field VEV when the corresponding mode $k_\star$ is leaving the Hubble patch during inflation.

Note that $\Delta\phi>0$ implies that the left 
hand side of the integration over momentum within an interval, $k_{e}<k<k_{\star}$, is also positive, where
$k_e$ represents corresponding momentum scale at the end of inflation.
 Individual integrals 
involving $\epsilon_V$ and $\eta_V$ are estimated in an Appendix, see Eqs.~(\ref{hj1}) and (\ref{hj2}).

In order to perform the momentum integration in the left hand side of Eq~(\ref{con4}), we have used the running of
 $r(k)$, which can be expressed as:
\be\label{con5}
 r(k)=r(k_{\star})\left(\frac{k}{k_{\star}}\right)^{a+\frac{b}{2}\ln\left(\frac{k}{k_{\star}}\right)
+\frac{c}{6}\ln^{2}\left(\frac{k}{k_{\star}}\right)+....}\,,
\ee
where 
\be
a=n_{T}-n_{S}+1,~~~b=\left(\alpha_{T}-\alpha_{S}\right),~~~c=\left(\kappa_{T}-\kappa_{S}\right)\,.
\ee
 These parameterisation characterises the spectral indices, $n_S,~n_T$, running 
of the spectral indices, $\alpha_S,~\alpha_T$, and running of the running of the spectral indices, $\kappa_S,~\kappa_T$. 
Here the subscripts, $(S,~T)$, represent the scalar and tensor modes. 
Now substituting the explicit form of the 
potential stated in Eq.~(\ref{rt1a}), we can evaluate the crucial integrals of the first and second slow-roll parameters ($\epsilon_{V},~\eta_{V}$)
appearing in the right hand side of Eq.~(\ref{con4}). For the details of the computation, see appendix.

In the present context the potential dependent slow-roll parameters:
($\epsilon_{V},~\eta_{V}$), satisfy the joint (Planck+WMAP-9) constraints, which imply that~\cite{Planck-infl}:
\be
\epsilon_V<10^{-2}\,,~~~5\times 10^{-3}<|\eta_{V}|<0.021\,,
\ee
for which the  inflationary potential is concave in nature.
In the next section, we will discuss model independent bounds on the coefficients ($V(\phi_\star),V^{\prime}(\phi_\star),\cdots$) for a generic sub-Planckian 
VEV inflationary setup, for which we will satisfy the joint constraints from: (Planck+WMAP-9+high~L+BICEP2), where $r(k_\star)=0$ is disfavoured at $7\sigma$ CL., see Eq.~(\ref{obscons}).


\section{Constraining the scale of inflation }

The number of
e-foldings, ${N}(k)$,  can be expressed as~\cite{Burgess:2005sb}:
%
\be\begin{array}{llll}\label{efold}
\displaystyle {N}(k) \approx  71.21 - \ln \left(\frac{k}{k_{\star}}\right)  
+  \frac{1}{4}\ln{\left( \frac{V_{\star}}{M^4_{P}}\right) }
+  \frac{1}{4}\ln{\left( \frac{V_{\star}}{\rho_{e}}\right) }  
+ \frac{1-3w_{int}}{12(1+w_{int})} 
\ln{\left(\frac{\rho_{rh}}{\rho_{e}} \right)},
\end{array}\ee
%
where $\rho_{e}$ is the energy density at the end of inflation, 
$\rho_{rh}$ is an energy scale during reheating, 
$k_{\star}=a_\star H_\star$ is the present Hubble scale, 
$V_{\star}$ corresponds to the potential energy when the relevant modes left the Hubble patch 
during inflation corresponding to the momentum scale $k_{\star}$, and $w_{int}$ characterises the effective equation of state 
parameter between the end of inflation, and the energy scale during reheating. 

Within the momentum interval,
$k_{e}<k<k_{\star}$, the corresponding number of e-foldings is given by, $\Delta {N} = {N}_{e}-{N}_{\star}$, as
\be\begin{array}{lll}\label{intnk}
    \displaystyle \Delta {N} =\ln\left(\frac{k_{\star}}{k_{e}}\right)\approx \ln\left(\frac{k_{\star}}{k_{e}}\right)
=\ln\left(\frac{a_{\star}}{a_{e}}\right)+\ln\left(\frac{H_{\star}}{H_{e}}\right)
\approx \ln\left(\frac{a_{\star}}{a_{e}}\right)+\frac{1}{2}\ln\left(\frac{V_{\star}}{V_{e}}\right)\,
   \end{array}\ee
where $(a_{\star},H_{\star})$ and $(a_{e}H_{e})$
represent the scale factor and the Hubble parameter at the pivot scale and end of inflation, and we have used the fact that 
$H^{2} \propto V$. We can estimate the contribution of the last term of the right hand side by using Eq~(\ref{rt1a}), and  as 
it  follows: $\ln\left({V_{\star}}/{V_{e}}\right)\approx \ln(1+\sqrt{2\epsilon_{V}}(\Delta \phi/M_p)(1+(\ll 1)))\approx 0$,
 where $(\Delta\phi/M_{p})<<1$, and$\sqrt{\epsilon_{V}} <<1$, consequently, Eq~(\ref{intnk}) reduces to:
\be\begin{array}{lll}\label{intnk1}
    \displaystyle \Delta {N} \approx \ln\left(\frac{k_{\star}}{k_{e}}\right)
\approx\ln\left(\frac{a_{\star}}{a_{e}}\right) \approx 17~{\rm efolds}\,.
   \end{array}\ee
Within the observed limit of Planck, i.e. $\Delta N\approx 17$, the slow-roll parameters, see  Eqs.~(\ref{hj1},~\ref{hj2}) of Appendix, show non-monotonic behaviour, 
where the corresponding scalar and tensor amplitude of the power spectrum remains almost unchanged.
Substituting the results obtained from Eq.~(\ref{hj1}) and Eq.~(\ref{hj2}) (see Appendix), and with the help of Eq.~(\ref{intnk}), 
up to the leading order, we obtain:
%
\be\begin{array}{llll}\label{con9}
    \displaystyle \sum^{\infty}_{n=0}{\bf G}_{n}\left(\frac{|\Delta\phi|}{M_p}\right)^{n}\approx\displaystyle\sqrt{\frac{r(k_{\star})}{8}}\times
\left|\frac{a}{4}-\frac{b}{16}+\frac{c}{48}-\frac{1}{2}\right|+\cdots\end{array}\ee
where we have used $(k_{e}/k_{\star})\approx \exp(-\Delta{N})=\exp(-17)\approx 4.13\times 10^{-8}$. 
 Here we will concentrate on $a,~b,~c \neq 0$, and  $a>>b>>c$ case is satisfied (for the details, see~\cite{Choudhury:2013iaa}) 
 by the joint constraints obtained from  (Planck+WMAP-9+high~L+BICEP2) data where $r(k_\star)=0$ is disfavoured at $7\sigma$ CL. 
 In Eq~(\ref{con9}) the series appearing in the left side of the above expression is convergent, since the expansion coefficients can be expressed as:
\be\begin{array}{lll}\label{expco}
\displaystyle {\bf G}_{n}=\left\{\begin{array}{ll}
                    \displaystyle  \left(1 +\underbrace{\sum^{\infty}_{m=0}{\bf A}_{m}
\left(\frac{\phi_{e}-\phi_{0}}{M_p}\right)^{m}}_{\ll 1} \right)\sim 1 &
 \mbox{ {\bf for} $n=1$}  \\ 
         \displaystyle  << 1 & \mbox{ {\bf for} $n\geq 2$}\,,
  \end{array}
\right. \end{array}\ee
%
and we have defined a new dimensionless 
binomial expansion co-efficient (${\bf A}_{m}$), defined as:
\be\label{cond}
{\bf A}_{m}=M^{m+2}_{p}\left[\left({\cal C}_{E}-\frac{1}{2}\right){\bf C}_{m}-{\cal C}_{E}{\bf D}_{m}\right]~~~~(\forall m=0,1,2,....)\,,
 \ee
which is obtained from the binomial series expansion  from the
leading order results of the slow-roll integrals stated in the appendix~\footnote{In Eq.~(\ref{cond}), and 
Eqs.~(\ref{hj1},~\ref{hj2}) (see Appendix), ${\bf C}_{p}$ and ${\bf D}_{q}$ are Planck suppressed dimensionful (mass dimension 
$\left[M^{-(m+2)}_{p}\right]$) binomial series expansion coefficients, which are expressed in terms of the generic model 
parameters $(V(\phi_\star),V^{\prime}(\phi_\star),\cdots)$ as presented
in Eq~(\ref{rt1a}).}.

Note that the expansion co-efficient ${\bf A}_{m}(\forall m)$ are suppressed by, 
$V(\phi_0)$, which is the leading order term in a generic expansion of the 
inflationary potential as shown in Eq~(\ref{rt1a}) (see Eq~(\ref{ceff}) in the appendix). 
We can expand the left side of Eq.~(\ref{con9}) 
in the powers of ${\Delta\phi}/{M_p}$, using the additional constraint 
$\Delta\phi<(\phi_{e}-\phi_{0})<M_p$, and we keep the leading order terms in $\Delta\phi/M_p$.


To the  first order approximation - we can 
neglect all the higher powers of $k_{e}/k_{\star}\approx {\cal O}(10^{-8})$ from the left hand side of Eq~(\ref{con9}), within  $17$ e-foldings of inflation.
Consequently, Eq.~(\ref{con9}) reduces to the following compact form for $r(k_\star)$:
%
\be\begin{array}{llll}\label{con10sd}
    \displaystyle \frac{9}{50}\sqrt{\frac{r(k_{\star})}{0.27}}\left|\frac{27}{1600}\left(\frac{r(k_{\star})}{0.27}\right)
-\frac{\eta_{V}(k_{\star})}{2}-\frac{1}{2}+\cdots\,\right|
\displaystyle \approx \frac{\left |\Delta\phi\right|}{M_p}\leq 1\,,
   \end{array}\ee
%
provided at the pivot scale, $k=k_{\star}>>k_{e}$, in this regime $\eta_{V}>>\left\{\epsilon^{2}_{V},\eta^{2}_{V},\xi^{2}_{V},\sigma^{3}_{V},\cdots\right\}$
approximation is valid.  Our expression, Eq.~(\ref{con10sd}),
shows that large value of $r(k_\star)$ can be obtained for models of inflation where
 inflation occurs below the Planck cut-off. Once the field excursion, $|\Delta\phi|/M_p$, and $\eta_{V}$, are known from any type of sub-Planckian inflationary 
setup, one can easily compute the tensor-to-scalar ratio by finding the roots~
\footnote{
The above expression, Eq~(\ref{con10sd}), is in the form of a simple algebraic (cubic) equation.  In order to find the roots of 
tensor-to-scalar ratio $r$ in terms of the field excursion $|\Delta\phi|/M_p$, one has to solve a cubic equation:
$x^{3}+Ux-W=0$,
where 
$$\displaystyle x:=\sqrt{\frac{r(k_{\star})}{0.27}},~~~U:=\frac{800}{27}\left(\eta_{V}+1\right),~~~W:=\frac{80000}{234}\frac{\left |\Delta\phi\right|}{M_p}$$
The three roots  $x_{1},x_{2},x_{3}$ are explicitly given by:
\bea\label{root1}
x_{1}&=&-\frac{\left(\frac{2}{3}\right)^{1/3} U}{Y}
+\frac{Y}{2^{1/3} 3^{2/3}},\\
x_{2}&=&\frac{\left(1+i \sqrt{3}\right) U}{2^{2/3} 3^{1/3} Y}
-\frac{\left(1-i \sqrt{3}\right) Y}{ 2^{4/3} 3^{2/3}},\\
x_{3}&=&\frac{\left(1-i \sqrt{3}\right) U}{2^{2/3} 3^{1/3} Y}
-\frac{\left(1+i \sqrt{3}\right) Y}{ 2^{4/3} 3^{2/3}}.
\eea
where the symbol, $Y=\left(9 W+\sqrt{3} \sqrt{4 U^3+27 W^2}\right)^{1/3}$. Here the complex roots $x_{2}$ and $x_{3}$ are physically redundant.
The only acceptable root is the real one, i.e. $x_{1}$.}.

Further note that our formulation will also hold true if inflation were to start at the hill-top, such as $\phi=0$. However in this case one would have 
to proceed similarly by expanding the potential around $\phi_0=0$, and then follow the algorithm we have provided here.

Now, it is also possible to recast $a(k),~b(k),~c(k)$, in terms of $r(k)$, and the slow roll parameters by using the 
relation, Eq.~(\ref{para 21e}) ( see Appendix ):
\be\begin{array}{lll}\label{apara}
    \displaystyle a(k_{\star})\approx \left[\frac{r(k_{\star})}{4}-2\eta_{V}(k_{\star})
+\cdots\right],\\
    \displaystyle b(k_{\star})\approx \left[16\epsilon^{2}_{V}(k_{\star})-12\epsilon_{V}(k_{\star})\eta_{V}(k_{\star})+2\xi^{2}_{V}(k_{\star})+\cdots\right],\\
\displaystyle c(k_{\star})\approx \left[-2\sigma^{3}_{V}+\cdots\right],
   \end{array}\ee
where $``\cdots''$ involve the higher order slow roll contributions, which are negligibly small in the leading order approximation.

The recent observations from (Planck+WAMP-9+high~L+BICEP2) have put an bound, $0.15\leq r(k_{\star})\leq 0.27$ at the pivot scale, $k_{\star}=0.002~Mpc^{-1}$, which can be expressed as an upper bound on $V(\phi_\star)$:
\begin{equation}\label{scale}
     V(\phi_{\star})=\frac{3}{2}P_{S}(k_{\star})r(k_{\star})\pi^{2}M^{4}_{p}\leq (2.40\times 10^{16}{\rm GeV})^{4}~\frac{r(k_{\star})}{0.27}.
   \end{equation}
The equivalent statement can be made in terms of the upper bound on the
numerical value of the Hubble parameter at the exit of the relevant modes:
\begin{equation}\label{hubinfla}
     H_{\star}\leq 1.38\times 10^{14}\times\sqrt{\frac{r(k_{\star})}{0.27}}~{\rm GeV}.
   \end{equation}
 Combining Eqs.~(\ref{con10sd}), (\ref{hubinfla}) and (\ref{scale}),
we can now obtain a closed relationships:
\begin{equation}\label{con13}
\frac{|\Delta\phi|}{M_p}  \leq  \frac{\sqrt{V_{\star}}}{(2.20\times 10^{-2}~M_p)^{2}}\left|\frac{V_{\star}}{(2.78\times 10^{-2}M_p)^{4}}-\frac{\eta_{V}(k_{\star})}{2}-\frac{1}{2} \right|\,.
\end{equation}

\begin{equation}\label{con13xc}
\frac{|\Delta\phi|}{M_p}  \leq  \frac{H_{\star}}{(2.79\times 10^{-4}~M_p)}\left|\frac{H^{2}_{\star}}{(1.99\times 10^{-7}~M^{2}_{p})}-\frac{\eta_{V}(k_{\star})}{2}-\frac{1}{2} \right|\,.
\end{equation}
where
$\eta_{V}>>\left\{\epsilon^{2}_{V},\eta^{2}_{V},\xi^{2}_{V},\sigma^{3}_{V},\cdots\right\}$ are satisfied. Similar expressions were derived in Ref.~\cite{Choudhury:2013iaa}
for {\it inflection-point} model of inflation, i.e. $V^{\prime\prime}(\phi_0)=0$ for (WMAP-9+Planck+High~L) datasets.
The above Eqs.~(\ref{con13},~\ref{con13xc}) are  bounds on $\Delta \phi$ for $\phi_0 < M_p$ and $\Delta \phi \ll M_p$.

Our conditions, Eqs.~(\ref{con10sd},~\ref{con13}), provide 
new constraints on model building for inflation within  particle theory, where the inflaton potential is always constructed 
within an effective field theory with a cut-off. Note that $|\eta_V(k_{\star})|> 0 $ can provide the largest contribution, 
in order to satisfy the (Planck+WAMP-9+high~L+BICEP2) bound on the tensor-to-scalar ratio within, $r=0.20^{+0.07}_{-0.05}$,
the shape of the potential has to be {\it concave}. Further applying this input in Eq~(\ref{scale}), one can get a preferred bound 
on the sub-Planckian VEV inflation, which lies within
 this tiny window:
  \be\label{gutr}2.07\times10^{16}~{\rm GeV}\leq\sqrt[4]{V_{\star}}\leq 2.40\times 10^{16}~{\rm GeV}.\ee
%

\begin{figure}[t]
\centering
\subfigure[$\epsilon_{V}$~vs~$\phi-\phi_{0}$]{
    \includegraphics[width=7.1cm, height=5.0cm] {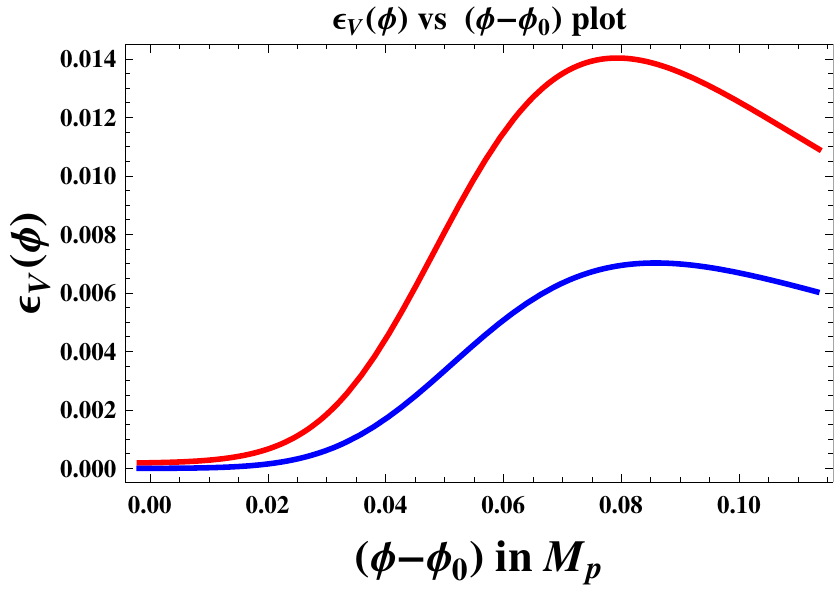}
    \label{fig:subfig1}
}
\subfigure[$|\eta_{V}|$~vs~$\phi-\phi_{0}$]{
    \includegraphics[width=7.1cm, height=5.0cm] {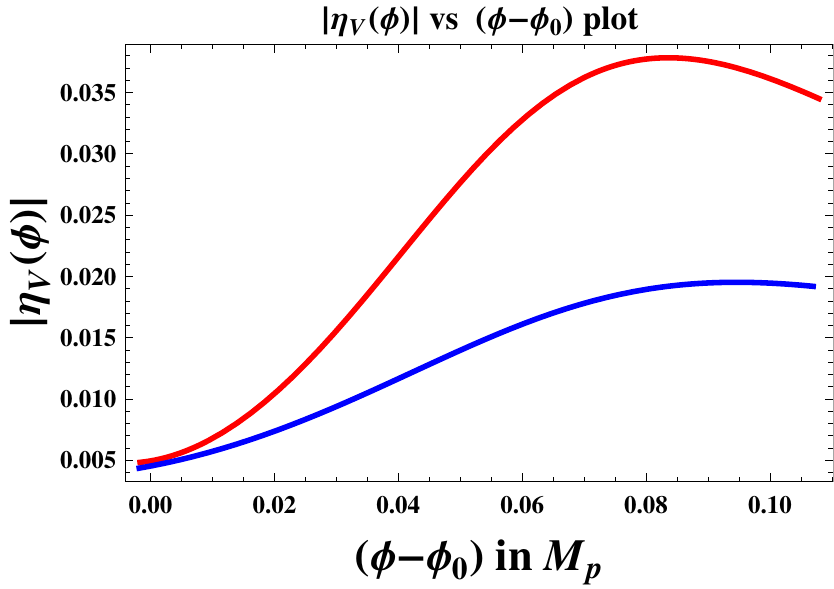}
    \label{fig:subfig2}
}
\subfigure[$|\xi^{2}_{V}|$~vs~$\phi-\phi_{0}$]{
    \includegraphics[width=7.1cm, height=5.0cm] {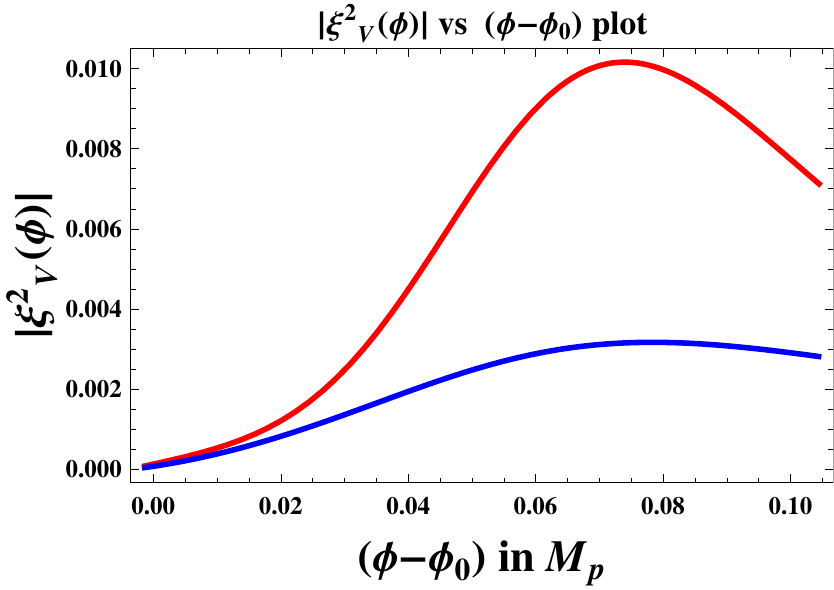}
    \label{fig:subfig3}
}
\subfigure[$|\sigma^{3}_{V}|$~vs~$\phi-\phi_{0}$]{
    \includegraphics[width=7.1cm, height=5.0cm] {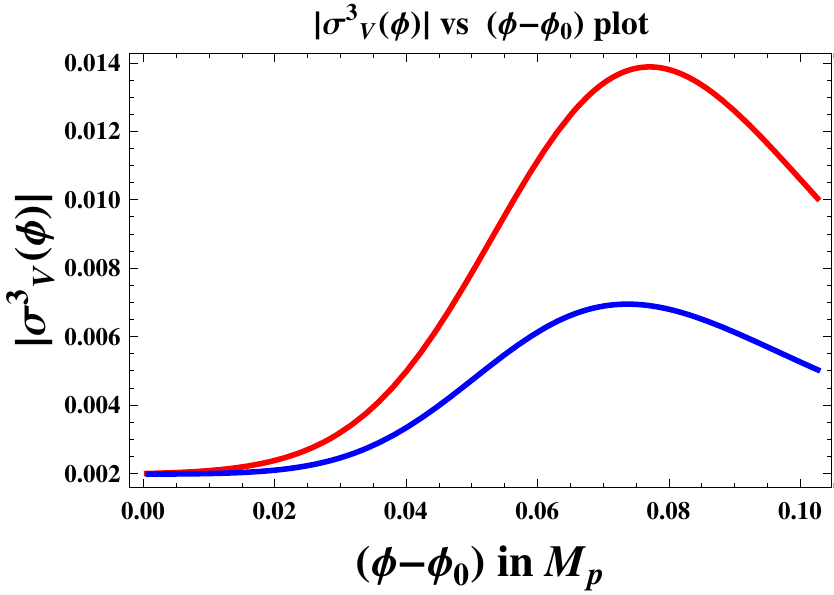}
    \label{fig:subfig4}
}
\caption[Optional caption for list of figures]{Non-monotonous evolution of the slow roll parameters are shown with respect to $\phi-\phi_0$. The upper and lower bounds are set by
Eqs.~(\ref{constraint6}-\ref{constraint10}).  
}
\label{fig3}
\end{figure}


\section{Reconstructing the potential }

Let us now discuss observational  constraints on the coefficients $(V(\phi_0),V^{'}(\phi_0),\cdots)$ of Eq~(\ref{rt10a}). Let us first write down 
$V(\phi_{\star}),V^{\prime}(\phi_{\star}),V^{\prime\prime}(\phi_{\star}),\cdots$ in terms of the inflationary observables ( see Appendix, 
Eqs.~(\ref{ps},~\ref{para 21c},~\ref{para 21e}) ):
\be\begin{array}{llll}\label{aq1}
\displaystyle V(\phi_{\star})= \frac{3}{2}P_{S}(k_{\star})r(k_{\star})\pi^{2}M^{4}_{p},\\
\displaystyle V^{'}(\phi_{\star})= \frac{3}{2}P_{S}(k_{\star})r(k_{\star})\pi^{2}\sqrt{\frac{r(k_{\star})}{8}}M^{3}_{p},\\
\displaystyle V^{''}(\phi_{\star})= \frac{3}{2}P_{S}(k_{\star})r(k_{\star})\pi^{2}\left(n_{S}(k_{\star})-1+\frac{3r(k_{\star})}{8}\right)M^{2}_{p},\\
\displaystyle V^{'''}(\phi_{\star})= \frac{3}{2}P_{S}(k_{\star})r(k_{\star})\pi^{2}\left[\sqrt{2r(k_{\star})}\left(n_{S}(k_{\star})-1+\frac{3r(k_{\star})}{8}\right)
\right.\\ \left.~~~~~~~~~~~~~~~~~~~~~~~~~~~~~~~~~~~~~~~~~~~~~~~~~~~~~~~~~~~~\displaystyle -\frac{1}{2}
\left(\frac{r(k_{\star})}{8}\right)^{\frac{3}{2}}-\alpha_{S}(k_{\star})\sqrt{\frac{2}{r(k_{\star})}}\right]M_{p},\\
\displaystyle V^{''''}(\phi_{\star})=12P_{S}(k_{\star})\pi^{2}\left\{\frac{\kappa_{S}(k_{\star})}{2}-
\frac{1}{2}\left(\frac{r(k_{\star})}{8}\right)^{2}\left(n_{S}(k_{\star})-1+\frac{3r(k_{\star})}{8}\right)\right.\\ \left.\displaystyle 
~~~~~~~~~~~~~~~~~~~~~~~~~~~~~~~~~~+12\left(\frac{r(k_{\star})}{8}\right)^{3}+r(k_{\star})\left(n_{S}(k_{\star})-1+\frac{3r(k_{\star})}{8}\right)^{2}
\right.\\ \left.\displaystyle 
~~~~~~~~~~~~~~~~~~~~~~~~~~~~~~~~~~+\left[\sqrt{2r(k_{\star})}\left(n_{S}(k_{\star})-1+\frac{3r(k_{\star})}{8}\right)
\right.\right.\\ \left.\left.~~~~~~~~~~~~~~~~~~~~~~~~~~~~~~~~~~~~~~~~~~~~~~~~~~~~~~~~~~~~\displaystyle -\frac{1}{2}
\left(\frac{r(k_{\star})}{8}\right)^{\frac{3}{2}}-\alpha_{S}(k_{\star})\frac{2}{r(k_{\star})}\right]\right.\\ \left.
\displaystyle 
~~~~~~~~~~~~~~~~~~~~~~~~~~~~~~~~~~\times \left[\sqrt{\frac{r(k_{\star})}{8}}\left(n_{S}(k_{\star})-1+\frac{3r(k_{\star})}{8}\right)
-6\left(\frac{r(k_{\star})}{8}\right)^{\frac{3}{2}}\right]\right\}
\end{array}\ee

Therefore, for any chosen sub-Planckian VEV of $\phi_{\star}$, we can obtain a matrix  equation characterising the coefficients:
 $V(\phi_0),~V^{\prime}(\phi_0),~V^{\prime\prime}(\phi_0),\cdots$:
\begin{equation}\label{mat1}\underbrace{\left(\begin{tabular}{cccccc}
 $1$ & $\vartheta_{\star}$ & $\frac{\vartheta^{2}_{\star}}{2}$& $\frac{\vartheta^{3}_{\star}}{6}$ & $\frac{\vartheta^{4}_{\star}}{24}$\\
$0$ & $1$ & $\vartheta_{\star}$& $\frac{\vartheta^{2}_{\star}}{2}$ & $\frac{\vartheta^{3}_{\star}}{6}$\\
$0$ & 0 & 1& $\vartheta_{\star}$ & $\frac{\vartheta^{2}_{\star}}{2}$\\
$0$ & $0$ & $0$& $1$ & $\vartheta_{\star}$\\
$0$ & $0$ & $0$& $0$ & $1$
      \end{tabular}\right)}
\left(\begin{tabular}{c}
      $V(\phi_{0})$\\
      $V^{'}(\phi_0)$\\
      $V^{''}(\phi_0)$\\
      $V^{'''}(\phi_0)$\\
      $V^{''''}(\phi_0)$\\
      \end{tabular}\right)
\begin{tabular}{c}
      \\
      =\\
      \\
      \end{tabular}
\left(\begin{tabular}{c}
      $V(\phi_{\star})$\\
      $V^{'}(\phi_{\star})$\\
      $V^{''}(\phi_{\star})$\\
     $V^{'''}(\phi_{\star})$\\
      $V^{''''}(\phi_{\star})$\\
      \end{tabular}\right)\,,
\end{equation}
where we have define $$\vartheta_{\star}:=(\phi_{\star}-\phi_0)\ll M_p.$$
The square matrix marked by the symbol $\underbrace{\cdots}$ is nonsingular,
since its determinant is nonzero, for which the matrix inversion technique is applicable in
the present context.
Finally, we get the following physical solution to the problem:
\begin{equation}\label{mat2}
\left(\begin{tabular}{c}
      $V(\phi_{0})$\\
      $V^{'}(\phi_0)$\\
      $V^{''}(\phi_0)$\\
      $V^{'''}(\phi_0)$\\
      $V^{''''}(\phi_0)$\\
      \end{tabular}\right)
\begin{tabular}{c}
      \\
      =\\
      \\
      \end{tabular}
\left(\begin{tabular}{cccccc}
 $1$ & $-\vartheta_{\star}$ & $\frac{\vartheta^{2}_{\star}}{2}$& $-\frac{\vartheta^{3}_{\star}}{6}$ & $\frac{\vartheta^{4}_{\star}}{24}$\\
$0$ & $1$ & $-\vartheta_{\star}$& $\frac{\vartheta^{2}_{\star}}{2}$ & $-\frac{\vartheta^{3}_{\star}}{6}$\\
$0$ & 0 & 1& $-\vartheta_{\star}$ & $\frac{\vartheta^{2}_{\star}}{2}$\\
$0$ & $0$ & $0$& $1$ & $-\vartheta_{\star}$\\
$0$ & $0$ & $0$& $0$ & $1$
      \end{tabular}\right)
\left(\begin{tabular}{c}
      $V(\phi_{\star})$\\
      $V^{'}(\phi_{\star})$\\
      $V^{''}(\phi_{\star})$\\
     $V^{'''}(\phi_{\star})$\\
      $V^{''''}(\phi_{\star})$\\
      \end{tabular}\right).
\end{equation}
For a model independent constraint on the shape of the potential, let us  fix the parameter, $\vartheta_{\star}\sim {\cal O}(10^{-2}M_{p})$, which is applicable 
to a large class of sub-Planckian inflationary models. This assumption holds good for high scale inflation, but within sub-Planckian cut-off~\footnote{ We will comment on 
our choice of $\vartheta\sim {\cal O}(10^{-2}M_p)$. In principle one can take $\vartheta \sim10^{-1}M_p -10^{-4}M_p$, or smaller values.}.

In order to satisfy the preferred bounds, see Eq.~(\ref{obscons}),
 the following model independent theoretical constraints 
on $V(\phi_{\star}),V^{'}(\phi_{\star}),\cdots$ have to be imposed:
\bea\label{constraint1}
   5.27\times 10^{-9}M^{4}_{p}\leq V(\phi_{\star})\leq 9.52\times 10^{-9}M^{4}_{p},\\  
   \label{constraint2}
   2.45\times 10^{-10}M^{3}_{p}\leq V^{'}(\phi_{\star})\leq 1.75\times 10^{-9}M^{3}_{p}, \\
\label{constraint3}
   4.82\times 10^{-11}M^{2}_{p}\leq V^{''}(\phi_{\star})\leq 6.51\times 10^{-10}M^{2}_{p}, \\
\label{constraint4}
   6.35\times 10^{-10}M_{p}\leq V^{'''}(\phi_{\star})\leq 7.56\times 10^{-10}M_{p}, \\
\label{constraint5}
   5.56\times 10^{-10}\leq V^{''''}(\phi_{\star})\leq 4.82\times 10^{-9},
   \eea
Now, substituting the above expression in Eq~(\ref{mat2}), we obtain  model independent constraints on the coeffiecients, $V(\phi_0),V^{\prime}(\phi_0),\cdots$:
\bea\label{constraint6}
   5.26\times 10^{-9}M^{4}_{p}\leq V(\phi_0)\leq 9.50\times 10^{-9}M^{4}_{p},\\  
   \label{constraint7}
   2.44\times 10^{-10}M^{3}_{p}\leq V^{'}(\phi_0)\leq 1.74\times 10^{-9}M^{3}_{p}, \\
\label{constraint8}
   4.19\times 10^{-11}M^{2}_{p}\leq V^{''}(\phi_0)\leq 6.44\times 10^{-10}M^{2}_{p}, \\
\label{constraint9}
   6.29\times 10^{-10}M_{p}\leq V^{'''}(\phi_0)\leq 7.08\times 10^{-10}M_{p}, \\
\label{constraint10}
   5.56\times 10^{-10}\leq V^{''''}(\phi_0)\leq 4.82\times 10^{-9},
   \eea
Consequently, the slow-roll parameters $(\epsilon_{V},\eta_{V},\xi^{2}_{V},\sigma^{3}_{V})$ are constrained by:
\bea\label{constraint6a}
   \epsilon_{V}\sim {\cal O}(0.10-1.69)\times 10^{-2},\\  
   \label{constraint7a}
   |\eta_{V}|\sim{\cal O}(9.14\times 10^{-3}-0.06), \\
\label{constraint8a}
   |\xi^{2}_{V}|\sim{\cal O}(5.60\times 10^{-3}-0.014), \\
\label{constraint9a}
   |\sigma^{3}_{V}|\sim {\cal O}(2.28\times 10^{-4}-0.017).
\eea
Further, by applying the joint constraints from (Planck+WMAP-9+high~L+BICEP2), we obtain the following model
 independent upper and lower bounds on  $|\Delta\phi|/M_p$ by using Eq~(\ref{con10sd}) or Eq~(\ref{con13}):  
\be\begin{array}{llll}\label{con10sdqx}
    \displaystyle 0.066 \leq\frac{\left |\Delta\phi\right|}{M_p}\,\leq 0.092.
   \end{array}\ee

Finally following the present analysis we get the following
 constraints on the tensor spectral tilt, $n_{T}$, running of the tensor spectral tilt, $\alpha_{T}$, the
running of the tensor-to-scalar ratio $n_{r}$ and running of the running of tensor spectral tilt $\kappa_{T}$ and tensor-to-scalar ratio $\kappa_{r}$ as:
\bea\label{wq1}
-0.019<n_{T}<-0.033,\\
\label{wq2}
-2.97\times 10^{-4}<\alpha_{T}<2.86\times 10^{-5},\\
\label{wq3}
2.28\times 10^{-4}<|n_{r}|<0.010 ,\\
\label{wq4}
-0.11\times 10^{-4}<\kappa_{T}<-3.58\times 10^{-4},\\
\label{wq5}
-5.25\times 10^{-3}<\kappa_{r}<-6.27\times 10^{-3},
\eea
In principle we could have fixed $\vartheta\sim {\cal O}(10^{-3}~M_{p}-10^{-4}~M_{p})$, in which case the numerical values 
of the coefficients $V(\phi_0),V^{\prime}(\phi_0),\cdots$ and the slow-roll parameters remain unaltered. Further, if we had set $\vartheta$ to a slightly
larger value, $\vartheta\sim {\cal O}(10^{-1}~M_{p})$, then the order of magnitude of the numerics would not change, but the numerical prefactors would slightly change.
For a varied range of $\vartheta\sim {\cal O}(10^{-1}M_p-10^{-4}M_p)$, our overall predictions presented in this paper remain robust.


\section{Sub-Planckian consistency relationships} 

Let us now provide the new set of consistency relationships {\it between slow roll parameters} for sub-Planckian models of inflation~\footnote{Just note that
for a super-Planckian models of inflation, where $\epsilon $ varies monotonically, we would obtain: $n_T=-r/8+\cdots$,~see Ref.~\cite{Lyth}.}:  
\begin{eqnarray}\label{wq6}n_{T}&=&-\frac{r}{8}\left(2-\frac{r}{8}-n_{S}\right)+\cdots,\\
\label{wq7}\alpha_{T}&=&\frac{dn_{T}}{d\ln k}=\frac{r}{8}\left(\frac{r}{8}+n_{S}-1\right)+\cdots,\\
\label{wq8} n_{r}&=&\frac{dr}{d\ln k}=\frac{16}{9}\left(n_{S}-1+\frac{3r}{4}\right)\left(2n_{S}-2+\frac{3r}{8}\right)+\cdots,\\
\label{wq9} \kappa_{T}&=&\frac{d^{2}n_{T}}{d\ln k^{2}}=\frac{2}{9}\left(n_{S}-1+\frac{3r}{4}\right)\left(2n_{S}-2+\frac{3r}{8}\right)\left(\frac{r}{8}+n_{S}-1\right)\\
&&~~~~~~~~~~~~~~~~~~+\frac{r}{8}\left[\alpha_{S}+\frac{2}{9}\left(n_{S}-1+\frac{3r}{4}\right)\left(2n_{S}-2+\frac{3r}{8}\right)\right]+\cdots,\\
\label{wq10} \kappa_{r}&=&\frac{d^{2}r}{d\ln k^{2}}\nonumber\\
&=&\frac{16}{9}\left(2n_{S}-2+\frac{3r}{8}\right)\left\{\alpha_{S}+\frac{4}{3}\left(n_{S}-1+\frac{3r}{4}\right)\left(2n_{S}-2+\frac{3r}{8}\right)\right\}\\
&&~~~~~~+\frac{16}{9}\left(n_{S}-1+\frac{3r}{4}\right)\left\{2\alpha_{S}+\frac{2}{3}\left(n_{S}-1+\frac{3r}{4}\right)\left(2n_{S}-2+\frac{3r}{8}\right)\right\}+\cdots.\nonumber
\end{eqnarray}
One can compare these relationships with respect to super-Planckian models of inflation where the slow roll parameters vary monotonically, see Re.~\cite{Lyth}.
Observationally, now one can differentiate sub versus super Planckian excursion models of inflation with the help of the above consistency relationships.
In particular, the slope of the tensor modes, see Eq.~(\ref{wq6}),  will play a crucial role in deciding the fate of the sub-primordial inflation in the early universe.

In Fig.~(\ref{fig3}), we have shown the evolution of $\epsilon_V,~|\eta_V|,~|\xi^2_V|,~|\sigma_V^3|$ (see Eqs.~(\ref{slpara})) with respect to 
$|\phi-\phi_0|$, the upper and lower bounds are given by Eqs.~(\ref{constraint6}-\ref{constraint10}). In particular, note that the evolution of $\epsilon_V$ is
non-monotonic for sub-Planckian inflation for  $0.15\leq r(k_\ast)\leq 0.27$, which is in stark contrast 
with the Lyth-bound for the super-Planckian models of inflation, where $\epsilon_V$ can evolve monotonically for polynomial potentials~\cite{Lyth}. In future 
the data would be  would be sufficiently good to compare the running of the gravitational tensor perturbations, $n_T$, for sub-vs-super-Planckian 
excursions of the inflaton.

\begin{figure}[t]
\centering
\subfigure[$P_{S}$~vs~$n_{S}$]{
    \includegraphics[width=7cm, height=5.0cm] {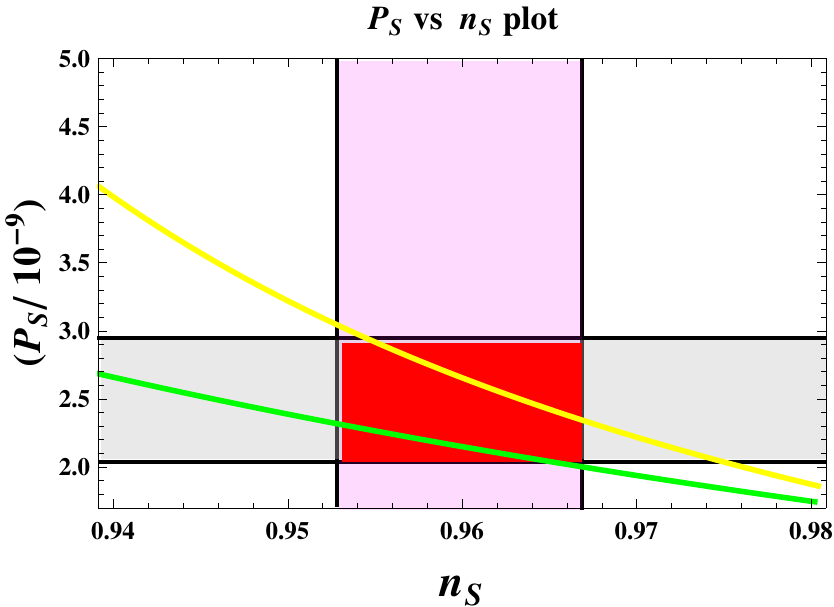}
    \label{fig:subfig5}
}
\subfigure[$P_{S}$~vs~$r_{0.002}$]{
    \includegraphics[width=7cm, height=5.0cm] {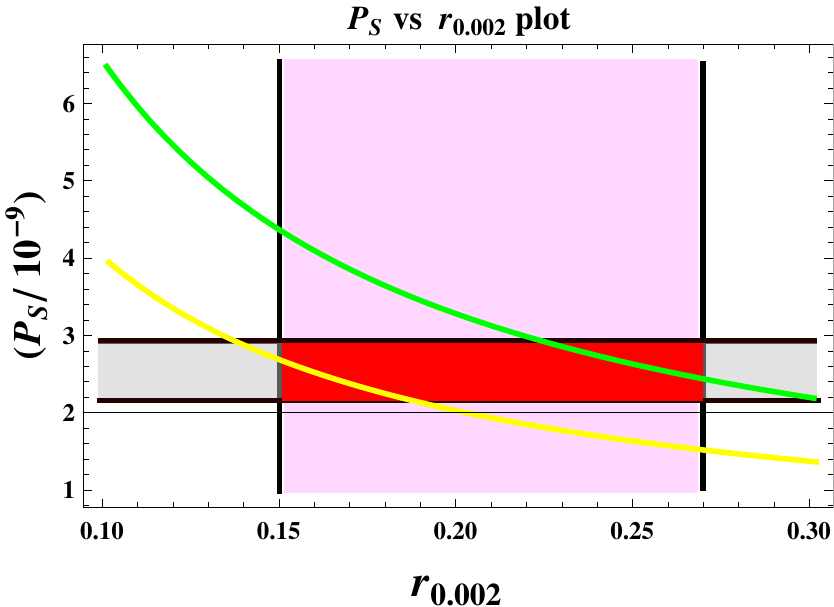}
    \label{fig:subfig6}
}
\caption[Optional caption for list of figures]{We have shown the variation of 
\subref{fig:subfig5} $P_{S}$~vs~$n_{S}$, and \subref{fig:subfig6} $P_{S}$~vs~$r_{0.002}$, for the pivot scale 
$k_{\star}=0.002~Mpc^{-1}$. The vertical patch shows the allowed region for BICEP2 data and the horizontal patch represents the allowed region by 
(Planck+WMAP-9+high-L) data for
both the cases. The overlapping {\it red} patch shows the allowed region by the joint constraints obtained from 
(Planck+WMAP-9+high-L+BICEP2).  The upper ({\it green}) and lower ({\it yellow}) bounds are set by
Eqs.~(\ref{constraint6}-\ref{constraint10}). 
}
\label{fig5}
\end{figure}

In Fig.~(\ref{fig:subfig5}) and Fig.~(\ref{fig:subfig6}), we have shown the variation of 
 $P_{S},~n_{S}$ and   $r_{0.002}$, at the pivot scale 
$k_{\star}=0.002~Mpc^{-1}$. The overlapping {\it red} patch shows the allowed region by the joint constraints obtained from (Planck+WMAP-9+high-L+BICEP2).
The upper ({\it green}) and lower ({\it yellow}) bounds are set by Eqs.~(\ref{constraint6}-\ref{constraint10}).

\begin{figure}[t]
{\centerline{\includegraphics[width=10.5cm, height=8.5cm] {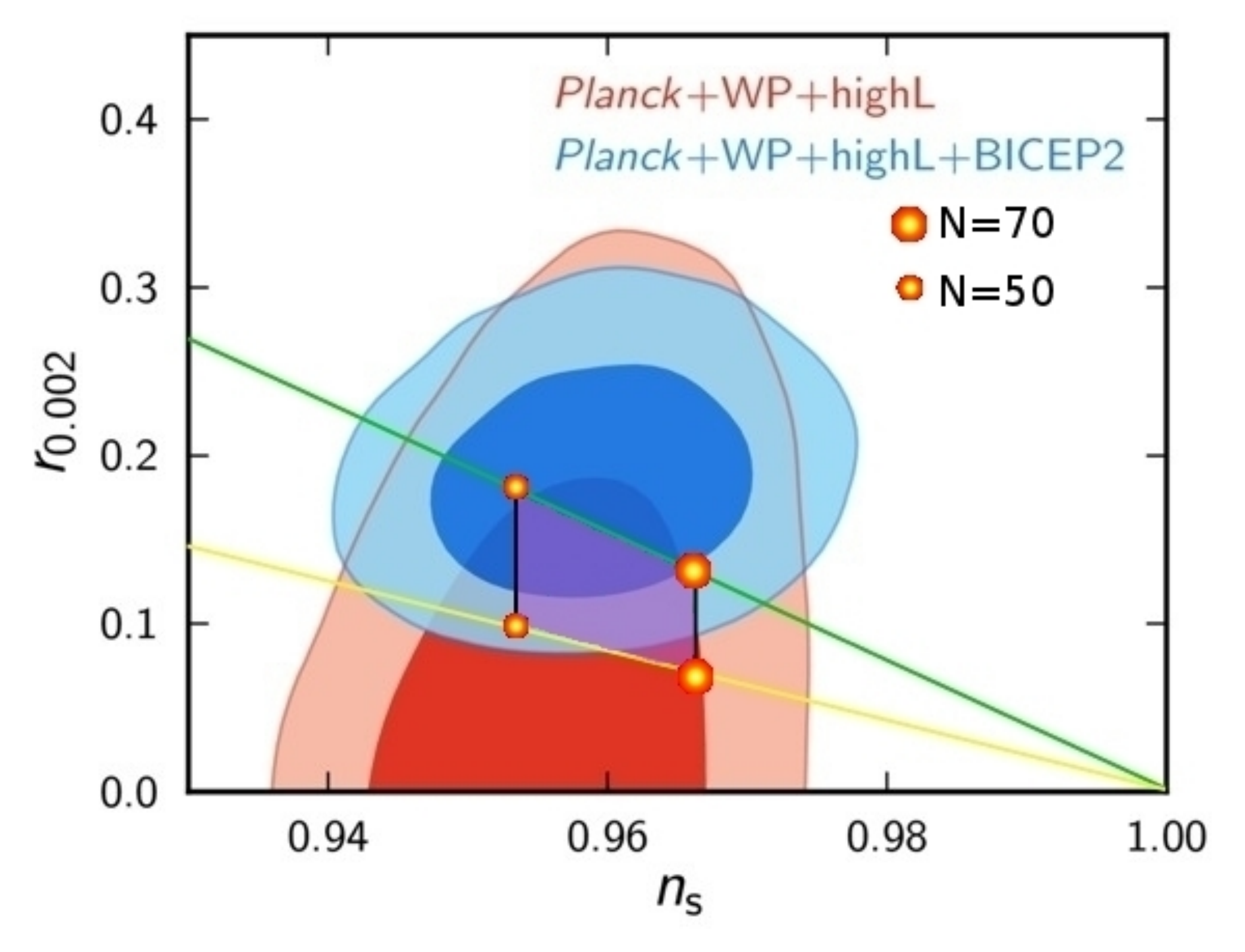}}}
\caption{We show the joint $1\sigma$ and $2\sigma$ CL. contours using
 (Planck+WMAP9+high~L) and (Planck+WMAP9+high~L+BICEP2) data for $r$ vs $n_{S}$ plot at
 the momentum pivot $k_{\star}\sim 0.002~{\rm Mpc}^{-1}$.
The small circle on the left corresponds to $N=50$, while the right big circle corresponds to $N=70$.
The allowed regions are shown by the shaded violet colour for $0.15<r_{\star}<0.27$ and  $0.952<n_{S}<0.967$.
The green and yellow lines are drawn for lower and upper bounds on the constraints derived in Eq~(\ref{constraint6}-\ref{constraint9a}), 
The vertical black coloured lines are drawn to show the bounded regions of the sub-Planckian inflationary model, along which
the number of e-foldings are fixed. } \label{fig1}
\end{figure}

In Fig~(\ref{fig1}), we have shown  $r$ vs $n_{S}$  at
the  pivot scale: $k_{\star}\sim 0.002~{\rm Mpc}^{-1}$.
The allowed region is shown by the shaded violet colour, for $0.15<r_{\star}<0.27$ and  $0.952<n_{S}<0.967$.
The green and yellow lines are drawn for lower and upper bound of the constraints derived in Eq~(\ref{constraint6}-\ref{constraint9a}).
We have used the relation between $n_{S}$ and $r_{\star}$ 
as mentioned in Eq~(\ref{para 21c},~\ref{para 21e}) in the appendix.
In Fig.~(\ref{fig2}), we have shown the plot of $r$ vs $\alpha_{S}=dn_{S}/d\ln k$ at
 the pivot scale: $k_{\star}\sim 0.05~{\rm Mpc}^{-1}$.

In Fig.~(\ref{fig:subfig7}-\ref{fig:subfig9}), we have depicted running of the tensor-to-scalar ratio: $n_{r}=dr/d\ln k$,
 running of the running of the tensor-to-scalar ratio: $\kappa_{r}=d^{2}r/d\ln k$,
running of the tensor spectral tilt: $\alpha_{T}=dn_{T}/d\ln k$,
 running of the running of tensor spectral tilt: $\kappa_{T}=d^{2}n_{T}/d\ln k$ vs scalar spectral tilt $n_{S}$.
Shaded violet colour region is the allowed region for (Planck+WMAP-9+high-L+BICEP2) which  constrain $n_{r},~\kappa_{r},~\alpha_{T}$ and $\kappa_{T}$
within the specified range mentioned in Eq~(\ref{wq2}) and Eq~(\ref{wq5}).
The green and yellow lines are drawn for lower and upper bound on the constraints derived in Eq~(\ref{constraint6}-\ref{constraint9a}), 
 We have used the relation between $\alpha_{T},~\kappa_{T}$ and $n_{S}$ 
as mentioned in Eq~(\ref{wq7}) and Eq~(\ref{wq9}).



\begin{figure}[t]
{\centerline{\includegraphics[width=8.5cm, height=7cm] {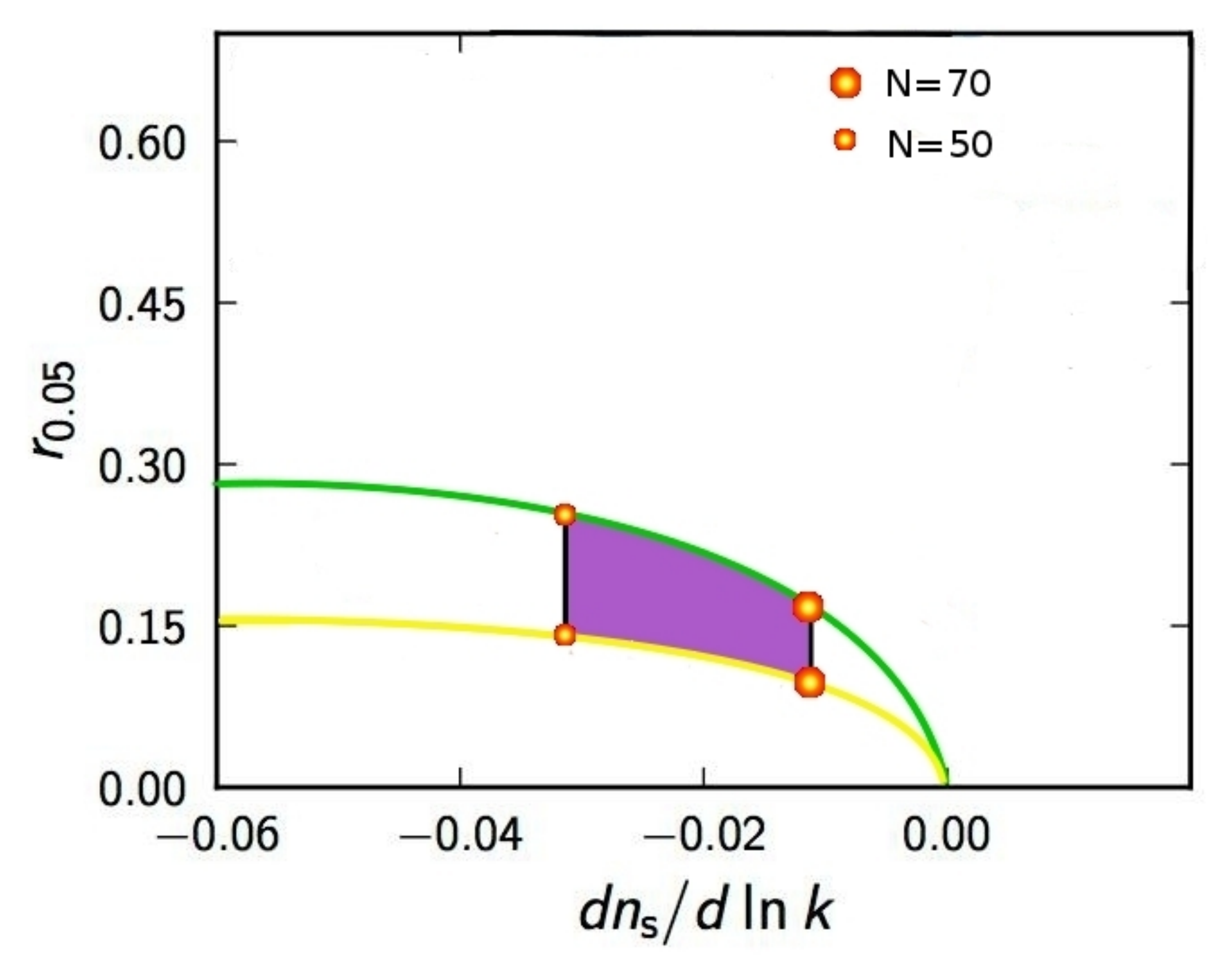}}}
\caption{We show  $r$ vs $\alpha_{S}=dn_{S}/d\ln k$ plot at
 the momentum pivot $k_{\star}\sim 0.05~{\rm Mpc}^{-1}$.
The small circle on the left corresponds to $N=50$, while the right big circle corresponds to $N=70$.
Shaded violet coloured regions are the allowed regions for $0.15<r_{\star}<0.27$ and  $-0.032<\alpha_{S}<-0.012$ specified by (Planck+WMAP-9+high-L+BICEP2).
The green and yellow lines are drawn for lower and upper bounds on the constraints derived in Eq~(\ref{constraint6}-\ref{constraint9a}), 
The vertical black coloured lines are drawn to show the bounded regions for the sub-Planckian inflationary model, along which
the number of e-foldings are fixed.  } \label{fig2}
\end{figure}



\begin{figure}[t]
\centering
\subfigure[$\alpha_{T}$~vs~$n_{S}$]{
    \includegraphics[width=7cm, height=5cm] {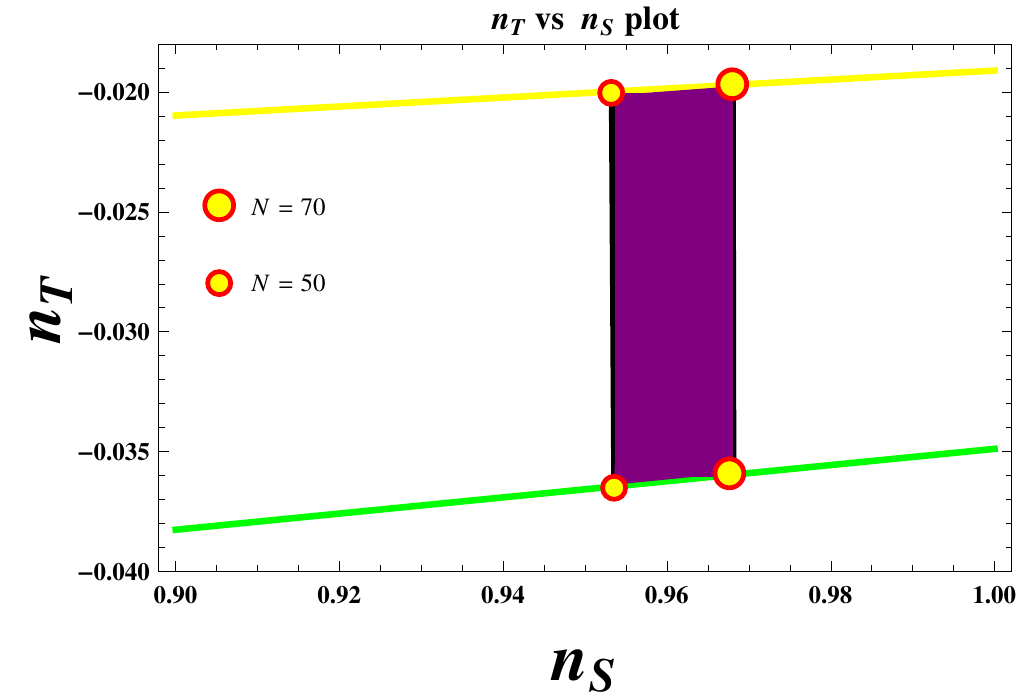}
    \label{fig:subfig7}
}
\subfigure[$\alpha_{T}$~vs~$n_{S}$]{
    \includegraphics[width=7cm, height=5cm] {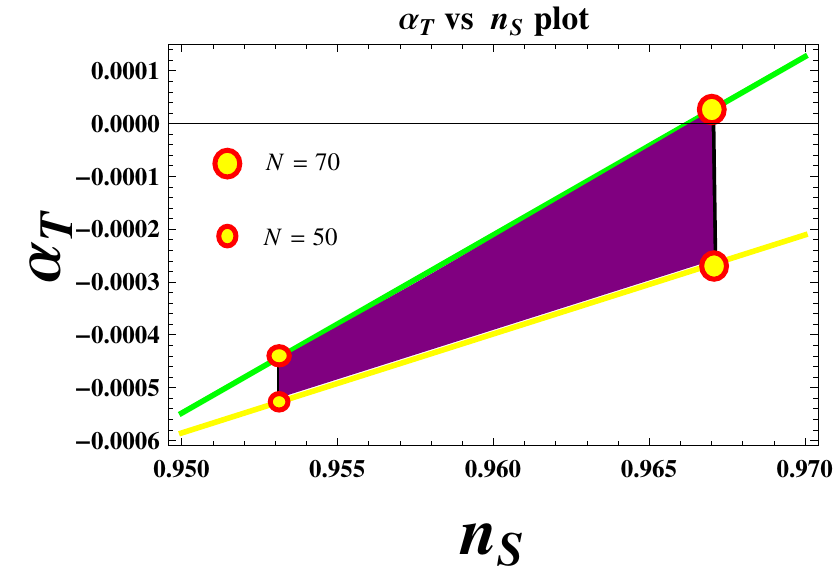}
    \label{fig:subfig8}
}
\subfigure[$\kappa_{T}$~vs~$n_{S}$]{
    \includegraphics[width=7cm, height=5cm] {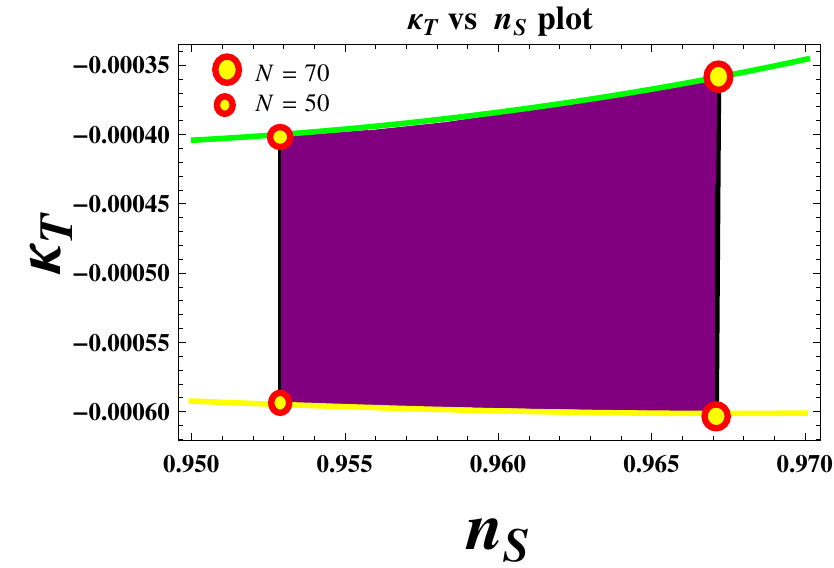}
    \label{fig:subfig9}
}
\caption[Optional caption for list of figures]{We show the \subref{fig:subfig7} tensor spectral tilt $n_{T}$, \subref{fig:subfig8} running of the tensor spectral tilt $\alpha_{T}=dn_{T}/d\ln k$,
\subref{fig:subfig9} running of the running of tensor spectral tilt $\kappa_{T}=d^{2}n_{T}/d\ln k$ vs scalar spectral tilt $n_{S}$ plot.
The small circle on the left corresponds to $N=50$, while the right big circle corresponds to $N=70$.
Shaded violet coloured regions are the allowed regions for (Planck+WMAP-9+high-L+BICEP2) which will constrain $\alpha_{T}$ and $\kappa_{T}$
within the specified ranges mentioned in Eq~(\ref{wq2}) and Eq~(\ref{wq5}).
The green and yellow lines are drawn for lower and upper bounds on the constraints derived in Eq~(\ref{constraint6}-\ref{constraint9a}), 
The vertical black coloured lines are drawn to show the bounded regions of the sub-Planckian inflationary model, along which
the number of e-foldings are fixed. 
}
\label{fig4}
\end{figure}



\section{Example of Inflection point inflation}

Now we impose, $V^{\prime\prime}(\phi_0)=0$, in order to study the {\it inflection point} scenario. The potential is given by~\footnote{The inflection point inflation 
has been studied in Refs.~\cite{Enqvist:2010vd}, with a constant potential energy density $V(\phi_0)$.}:
\be\label{rt1a}
V(\phi)=V(\phi_0)+V^{\prime}(\phi_0)(\phi-\phi_{0})+\frac{V^{\prime\prime\prime}(\phi_0)}{6}(\phi-\phi_{0})^{3}
+\frac{V^{\prime\prime\prime\prime}(\phi_0)}{24}(\phi-\phi_{0})^{4}+\cdots\,,
\ee
   %
We can express  $V(\phi_{\star}),V^{\prime}(\phi_{\star}),\cdots$ in terms of $\phi_0$ for a sub-Planckian regime as:
\bea\label{ws1}
V(\phi_{\star})&=&V(\phi_0)+\vartheta_{\star}V^{'}(\phi_0)+\frac{\vartheta^{3}_{\star}}{6}V^{'''}(\phi_0)+\frac{\vartheta^{4}_{\star}}{24}V^{''''}(\phi_0),\nonumber \\
V^{'}(\phi_{\star})&=&V^{'}(\phi_0)+\frac{\vartheta^{2}_{\star}}{2}V^{'''}(\phi_0)+\frac{\vartheta^{3}_{\star}}{6}V^{''''}(\phi_0),\nonumber \\
V^{''}(\phi_{\star})&=&\vartheta_{\star}V^{'''}(\phi_0)+\frac{\vartheta^{2}_{\star}}{2}V^{''''}(\phi_0),\nonumber \\
V^{'''}(\phi_{\star})&=&V^{'''}(\phi_0)+\vartheta_{\star}V^{''''}(\phi_0),\nonumber \\
V^{''''}(\phi_{\star})&=&V^{''''}(\phi_0).
\eea
Using Eq~(\ref{ws1}),  we obtain a {\it particular} solution for the coefficients: $V(\phi_0),~V^{\prime}(\phi_0),\cdots$, which can be written as
~\footnote{There will be in general $2$ solutions around an inflection point, here we will provide one of the two solutions which is the most interesting 
one for the general case of study.}: 
\bea\label{ws2}
V(\phi_0)&=&V(\phi_{\star})-\vartheta_{\star}V^{'}(\phi_{\star})+\frac{\vartheta^{3}_{\star}}{3}V^{'''}(\phi_{\star})-\frac{5\vartheta^{4}_{\star}}{24}V^{''''}(\phi_{\star}),\nonumber \\
V^{'}(\phi_0)&=&V^{'}(\phi_{\star})-\frac{\vartheta^{2}_{\star}}{2}V^{'''}(\phi_{\star})+\frac{\vartheta^{3}_{\star}}{3}V^{''''}(\phi_{\star}),\nonumber \\
V^{'''}(\phi_0)&=&V^{'''}(\phi_{\star})-\vartheta_{\star}V^{''''}(\phi_0),\nonumber \\
V^{''''}(\phi_0)&=&V^{''''}(\phi_{\star}).
\eea
Now using the bound on $V(\phi_{\star}),~V^{\prime}(\phi_{\star}),\cdots$ as mentioned in Eqs.~(\ref{constraint1}-\ref{constraint5}), we can 
obtain the following constraints on the coefficients of $V(\phi_0),V^{'}(\phi_0),\cdots$:
 \bea
\label{constraint11}
   5.26\times 10^{-9}M^{4}_{p}\leq V(\phi_0)\leq 9.50\times 10^{-9}M^{4}_{p},\\  
   \label{constraint12}
   2.44\times 10^{-10}M^{3}_{p}\leq V^{'}(\phi_0)\leq 1.74\times 10^{-9}M^{3}_{p}, \\
\label{constraint13}
   6.29\times 10^{-10}M_{p}\leq V^{'''}(\phi_0)\leq 7.08\times 10^{-10}M_{p}, \\
\label{constraint14}
   5.56\times 10^{-10}\leq V^{''''}(\phi_0)\leq 4.82\times 10^{-9},
   \eea
We can compare these results with that of the constraints mentioned in Eq~(\ref{constraint6}-\ref{constraint10}) for a generic sub-Planckian inflationary 
setup for $\vartheta_\star =\phi_\star-\phi_0 \leq10^{-2}M_p$ for a sub-Planckian VEV model of inflation. We find a very nice agreement which testifies 
the power of a model independent reconstruction of the potential.



 \section{Summary}

In this paper we have obtained the most general expressions for $r(k_\star),~|\Delta \phi|/M_p$ in terms of $V(\phi_\star)$ and $H(k_\star)$ for a sub-Plackian excursion
of inflation by taking into account of higher order slow roll conditions. In order to satisfy the observational data from (Planck+WMAP-9+high~L+BICEP2), see Eq.~(\ref{obscons}), one requires a non-monotonic evolution of $\epsilon_V$ parameter if one wants to build a model of inflation with a sub-Planckian VEV, 
as pointed out in Refs.~\cite{BenDayan:2009kv,Hotchkiss:2011gz}. In this paper we have reconstructed the potential 
around $\phi_0$ for $V(\phi_0),~V^{\prime}(\phi_0),~V^{\prime\prime}(\phi_0),~V^{\prime\prime\prime}(\phi_0)$ and $V^{\prime\prime\prime\prime}(\phi_0)$.
In order to satisfy the current observational constraints, $0.15<r_{\star}<0.27$,  we have found the scale of inflation to be within:
$2.07\times10^{16}~{\rm GeV}\leq\sqrt[4]{V_{\star}}\leq 2.40\times 10^{16}~{\rm GeV}$, for the VEV of inflaton varying within: $ 0.066 \leq\frac{\left |\Delta\phi\right|}{M_p}\,\leq 0.092$. 

We have also estimated a new set of {\it consistency relationships} for sub-Planckian model of inflation. One particular discriminator between sub-vs-super 
Planckian field excursion is the consistency relationships, in particular the slope of the tensor modes, i.e. $n_T=-(r/8)(2-(r/8)-n_s)< 0$ for $n_s=0.96$, 
see Eq.~(\ref{wq6}). This is in contrast with the super-Planckian excursion models of inflation, where typically one would expect $n_T=-(r/8)$. Furthermore, if the
data could be refined to constrain $\alpha_T$, then this would really seal the status of sub-vs-super Planckian models of inflation.
We have also found that  choice of  $\vartheta$ is rather insensitive to constraining the model parameters.
 
Typically, in particle physics, the nature and shape of the potential will not {\it just} be a single monomial. In principle the potential could contain quadratic, cubic and quartic interactions for a renormalizable theory, or even higher order non-renormalizable terms arising from integrating out the heavy degrees of freedom, see~\cite{Mazumdar:2010sa}. In this respect our results are important for reconstructing a particle physics motivated model of inflation in a successful way.


\section*{Acknowledgments}
SC thanks Council of Scientific and
Industrial Research, India for financial support through Senior
Research Fellowship (Grant No. 09/093(0132)/2010). AM is supported 
by the Lancaster-Manchester-Sheffield Consortium for Fundamental Physics under STFC grant ST/J000418/1.

\section*{Appendix}

The expressions for the slow-roll parameters ($\epsilon_{V},\eta_{V},\xi^{2}_{V},\sigma^{3}_{V}$) can be expressed as:
\be\begin{array}{llll}\label{slpara}
\epsilon_{V}\approx\frac{M^{2}_{p}}{2V(\phi_0)^{2}}\left[V^{'}(\phi_0)+V^{''}(\phi_0)\left(\phi-\phi_0\right)+\frac{V^{'''}(\phi_0)}{2}\left(\phi-\phi_0\right)^{2}
+\frac{V^{''''}(\phi_0)}{6}\left(\phi-\phi_0\right)^3+\cdots\right]^{2},\\
\eta_{V}\approx\frac{M^{2}_{p}}{V(\phi_0)}\left[V^{''}(\phi_0)+V^{'''}(\phi_0)\left(\phi-\phi_0\right)+\frac{V^{''''}(\phi_0)}{2}\left(\phi-\phi_0\right)^2+\cdots\right],\\
\xi^{2}_{V}\approx\frac{M^{4}_{p}}{V(\phi_0)^{2}}\left[V^{'''}(\phi_0)V^{'}(\phi_0)+\left(V^{''''}(\phi_0) V^{'}(\phi_0)+V^{'''}(\phi_0)V^{''}(\phi_0)\right)
\left(\phi-\phi_0\right)\right. \\ \left. ~~~~~~~~~~~~~~~~~~~~~~~~~~~~~~~~~~~~~~~~+\left(V^{''''}(\phi_0)V^{''}(\phi_0)+\frac{V^{'''}(\phi_0)^{2}}{2}\right)
\left(\phi-\phi_0\right)^{2}+\cdots\right],\\
\sigma^{3}_{V}\approx\frac{V^{''''}(\phi_0) M^{6}_{p}}{V(\phi_0)^{3}}\left[V^{'}(\phi_0)+V^{''}(\phi_0)\left(\phi-\phi_0\right)+\frac{V^{'''}(\phi_0)}{2}\left(\phi-\phi_0\right)^{2}
+\frac{V^{''''}(\phi_0)}{6}\left(\phi-\phi_0\right)^3+\cdots\right]^{2}.
\end{array}\ee
Furthermore, the inflationary observables, i.e. the amplitude of scalar and tensor power spectrum $(P_{S},P_{T})$, spectral tilt $(n_{S},n_{T})$, and tensor-to-scalar 
ratio $(r_{\star})$ at the pivot scale $k_{\star}$ can be expressed as:  
\be\begin{array}{lll}\label{ps}
\displaystyle P_{S}(k_{\star}) 
=\left[1-(2{\cal C}_{E}+1)\epsilon_{V}+{\cal C}_{E}\eta_{V}\right]^{2}\frac{V}{24\pi^{2}M^{4}_{p}\epsilon_{V}}\\
\displaystyle ~~~~~~~~~\approx\frac{V(\phi_0)^{2}}{12\pi^{2}M^{6}_{p}V^{'}(\phi_0)^{2}}\left[V(\phi_0)+V^{'}(\phi_0)(\phi_{\star}-\phi_{0})+\frac{V^{''}(\phi_0)}{2}(\phi_{\star}-\phi_{0})^{2}
+\frac{V^{'''}(\phi_0)}{6}(\phi_{\star}-\phi_{0})^{3}\right.\\ \left. \displaystyle ~~~~~~~~~~~~~~~~~~~~~~~~~~~~~+\frac{V^{''''}(\phi_0)}{24}(\phi_{\star}-\phi_{0})^{4}+\cdots\,\right]
\left[1-(2{\cal C}_{E}+1)\frac{V^{'}(\phi_0)^{2}M^{2}_{p}}{2V(\phi_0)^{2}}+{\cal C}_{E}\frac{M^{2}_{p} V^{''}(\phi_0)}{V(\phi_0)}\right]^{2},\\
\end{array}\ee
\be\begin{array}{lll}\label{pT}
\displaystyle  P_{T}(k_{\star})=\left[1-({\cal C}_{E}+1)\epsilon_{V}\right]^{2}\frac{2V}{3\pi^{2}M^{4}_{p}}\\
\displaystyle ~~~~~~~~~\approx\frac{2}{3\pi^{2}M^{4}_{p}}\left[V(\phi_0)+V^{'}(\phi_0)(\phi_{\star}-\phi_{0})+\frac{V^{''}(\phi_0)}{2}(\phi_{\star}-\phi_{0})^{2}
+\frac{V^{'''}(\phi_0)}{6}(\phi_{\star}-\phi_{0})^{3}\right.\\ \left. \displaystyle ~~~~~~~~~~~~~~~~~~~~~~~~~~~~~+\frac{V^{''''}(\phi_0)}{24}(\phi_{\star}-\phi_{0})^{4}+\cdots\,\right]
\left[1-({\cal C}_{E}+1)\frac{V^{'}(\phi_0)^{2}M^{2}_{p}}{2V(\phi_0)^{2}}\right]^{2},\\
\end{array}\ee
\be\begin{array}{lllll}\label{para 21c} \displaystyle  n_{S}-1
 \approx (2\eta_{V}
-6\epsilon_{V})+\cdots\\
~~~~~~~~~\displaystyle =M^{2}_{p}\left[\left(\frac{2V^{''}(\phi_0)}{V(\phi_0)}-\frac{V^{'}(\phi_0)^{2}}{2V(\phi_0)^{2}}\right)
+\left(\frac{2V^{'''}(\phi_0)}{V(\phi_0)}-\frac{2V^{'}(\phi_0)V^{''}(\phi_0)}{V(\phi_0)^{2}}\right)\left(\phi_{\star}-\phi_0\right)+\cdots\right],\end{array}\ee
\be\begin{array}{lllll}\label{para 21cc} \displaystyle  n_{T}
 \approx -2\epsilon_{V}+\cdots=-\frac{M^{2}_{p}}{V(\phi_0)^{2}}\left[V^{'}(\phi_0)+V^{''}(\phi_0)\left(\phi_{\star}-\phi_0\right)+\cdots\right]^{2},\end{array}\ee
\be\begin{array}{lllll}\label{para 21e} \displaystyle  r(k_{\star})\approx16\epsilon_{V}\left[1+2{\cal C}_{E}(\epsilon_{V}-\eta_{V})\right]+\cdots\\
~~~~~~~\displaystyle=\frac{8M^{2}_{p}}{V(\phi_0)^{2}}\left[V^{'}(\phi_0)+V^{''}(\phi_0)\left(\phi_{\star}-\phi_0\right)+\cdots\right]^{2}
\left[1+2{\cal C}_{E}M^{2}_{p}\left(\frac{V^{'}(\phi_0)^{2}}{2V(\phi_0)^{2}}-\frac{V^{''}(\phi_0)}{V(\phi_0)}\right)\right]
\end{array}\ee

\be\begin{array}{lllll}\label{para 21f}  \displaystyle \alpha_{S}(k_{\star})=\left(\frac{dn_{S}}{d\ln k}\right)_{\star}
\approx\left(16\eta_{V}\epsilon_{V}-24\epsilon^{2}_{V}-2\xi^{2}_{V}\right)+\cdots,\\
~~~~~~~~~\displaystyle =\frac{8M^{4}_{p}}{V(\phi_0)^3}\left[V^{'}(\phi_0)+V^{''}(\phi_0)\left(\phi_{\star}-\phi_0\right)+\cdots\right]^{2}
\left[V^{''}(\phi_0)+V^{'''}(\phi_0)\left(\phi_{\star}-\phi_0\right)+\cdots\right]\\
~~~~~~~~~~~~~~~~~~~~~~~~~~~~~~~~~~~~~~~~~~~~~~~\displaystyle -\frac{6M^{4}_{p}}{V(\phi_0)^{4}}\left[V^{'}(\phi_0)+V^{''}(\phi_0)\left(\phi_{\star}-\phi_0\right)
\cdots\right]^{2}\\
\displaystyle ~~~~~~~~~~~~~~~~~~~-\frac{2M^{4}_{p}}{V(\phi_0)^{2}}\left[V^{'''}(\phi_0)V^{'}(\phi_0)+\left(V^{''''}(\phi_0) V^{'}(\phi_0)+V^{'''}(\phi_0)V^{''}(\phi_0)\right)
\left(\phi_{\star}-\phi_0\right)\right. \\ \left. ~~~~~~~~~~~~~~~~~~~~~~~~~~~~~~~~~~~~~~~~+\left(V^{''''}(\phi_0)V^{''}(\phi_0)+\frac{V^{'''}(\phi_0)^{2}}{2}\right)
\left(\phi_{\star}-\phi_0\right)^{2}+\cdots\right]
\end{array}\ee

\be\begin{array}{lllll}\label{para 21h} \displaystyle   \kappa_{S}(k_{\star})=\left(\frac{d^{2}n_{S}}{d\ln k^{2}}\right)_{\star}\approx192\epsilon^{2}_{V}\eta_{V}-192\epsilon^{3}_{V}+2\sigma^{3}_{V}
-24\epsilon_{V}\xi^{2}_{V}+2\eta_{V}\xi^{2}_{V}-32\eta^{2}_{V}\epsilon_{V}+\cdots\\
~~~~~~~~~\displaystyle =\frac{48M^{6}_{p}V^{'}(\phi_0)^4V^{''}(\phi_0)}{V(\phi_0)^5}-\frac{24M^{6}_{p}V^{'}(\phi_0)^6}{V(\phi_0)^6}
-\frac{12M^{6}_{p}V^{'}(\phi_0)^3V^{'''}(\phi_0)}{V(\phi_0)^4}\\
~~~~~~~~~~~~~~~~~~\displaystyle +\frac{2M^{6}_{p}V^{'}(\phi_0)V^{''}(\phi_0)V^{'''}(\phi_0)}{V(\phi_0)^3}
-\frac{16M^{6}_{p}V^{'}(\phi_0)^2V^{''}(\phi_0)^2}{V(\phi_0)^4}\\
~~~~~~~~~~~~~~~~~~~~~~~~~~~~~~~~~~~~~~~~~~~~~~~~~\displaystyle +\frac{2M^{6}_{p}V^{'}(\phi_0)^2V^{''''}(\phi_0)}{V(\phi_0)^3}+\cdots\end{array}\ee

\be\begin{array}{lllll}\label{para 21red} \displaystyle   n_{r}(k_{\star})=\left(\frac{dr}{d\ln k}\right)_{\star}\approx
32\epsilon_{V}\eta_{V}-64\epsilon^{2}_{V}+\cdots\\
~~~~~~~~~\displaystyle =16M^{4}_{p}\left(\frac{V^{'}(\phi_0)^2V^{"}(\phi_0)}{V(\phi_0)^3}-\frac{V^{'}(\phi_0)^4}{V(\phi_0)^4}\right)+\cdots\end{array}\ee
The crucial integrals of the first and second slow-roll parameters ($\epsilon_{V},~\eta_{V}$)
appearing in the right hand side of Eq.~(\ref{con4}),
 which can be written up to the leading order as: 
%
\be\begin{array}{llll}\label{hj1}
    \displaystyle \int^{{\phi}_{\star}}_{{\phi}_{e}}d {\phi}~\epsilon_{V}
\approx\frac{1}{2}\sum^{\infty}_{p=0} \frac{M^{p+2}_{p}~{\bf C}_{p}}{(p+1)}\left(\frac{\phi_{e}-\phi_{0}}{M_p}\right)^{p+1}
\left\{\left(1+\frac{\Delta\phi}{M_p}\left(\frac{\phi_{e}-\phi_{0}}{M_p}\right)^{-1}\right)^{p+1}-
1\right \}+\cdots\,
   \end{array}\ee
 \be\begin{array}{llll}\label{hj2}
    \displaystyle \int^{{\phi}_{\star}}_{{\phi}_{e}}d {\phi}~\eta_{V}
\approx \sum^{\infty}_{q=0} \frac{M^{q+2}_{p}~{\bf D}_{q}}{(q+1)}\left(\frac{\phi_{e}-\phi_{0}}{M_p}\right)^{q+1}
\left\{\left(1+\frac{\Delta\phi}{M_p}\left(\frac{\phi_{e}-\phi_{0}}{M_p}\right)^{-1}\right)^{q+1}-
1\right \}+\cdots\,
   \end{array}\ee
%
where we have used the  $(\phi-\phi_{0})<M_p$  (including at $\phi=\phi_{\star}$ and $\phi=\phi_{e}$) around $\phi_0$.
The leading order dimensionful Planck scale suppressed expansion co-efficients (${\bf C}_{p}$) and (${\bf D}_{q}$)
 are given in terms of the model parameters $(V(\phi_0),V^{'}(\phi_0),\cdots)$, which determine the
hight and shape of the potential in terms of the model parameters as:
%
\be\begin{array}{llll}\label{ceff}
    \displaystyle {\bf C}_{0}=\frac{V^{'}(\phi_0)^{2}}{V(\phi_0)^{2}},~~{\bf C}_{1}=\frac{2V^{''}(\phi_0)V^{'}
(\phi_0)}{V(\phi_0)^{2}}-\frac{2V^{'}(\phi_0)^{3}}{V(\phi_0)^{3}},\\
\displaystyle {\bf C}_{2}=\frac{V^{''}(\phi_0)^{2}}{V(\phi_0)^{2}}-\frac{5V^{'}(\phi_0)^{2}V^{''}(\phi_0)}{V(\phi_0)^{3}}+\frac{V^{'}(\phi_0)V^{'''}(\phi_0)}{V(\phi_0)^{2}},\\
\displaystyle {\bf C}_{3}=\frac{V^{'}(\phi_0)V^{''''}(\phi_0)}{3V(\phi_0)^{2}}-\frac{7V^{'}(\phi_0)^{2}V^{'''}(\phi_0)}{3V(\phi_0)^{3}}+\frac{V^{''}(\phi_0)V^{'''}(\phi_0)}{V(\phi_0)^{2}}
\\~~~~~~~~~~~~~~~~~~~~~~~~~~~~~~~~~~~~~~\displaystyle-\frac{4V^{'}(\phi_0)V^{''}(\phi_0)^{2}}{V(\phi_0)^{3}}+\frac{9V^{'}(\phi_0)^{3}V^{''}(\phi_0)}{V(\phi_0)^{4}},\\
\displaystyle \cdots \cdots \cdots \cdots \cdots \cdots\cdots \cdots \cdots \cdots \cdots \cdots\cdots \cdots \cdots\\
    \displaystyle {\bf D}_{0}=\frac{V^{''}(\phi_0)}{V(\phi_0)},~~{\bf D}_{1}=\frac{V^{'''}(\phi_0)}{V(\phi_0)}-\frac{V^{'}(\phi_0)V^{''}(\phi_0)}{V(\phi_0)^{2}},\\
\displaystyle {\bf D}_{2}=\frac{V^{''''}(\phi_0)}{2V(\phi_0)}-\frac{V^{'}(\phi_0)V^{'''}(\phi_0)}{V(\phi_0)^{2}}-\frac{V^{''}(\phi_0)}{V(\phi_0)^{2}}+\frac{V^{''}(\phi_0) V^{'}(\phi_0)^{2}}{V(\phi_0)^{3}},\\
\displaystyle {\bf D}_{3}=\frac{4V^{'}(\phi_0)\delta^{2}}{V(\phi_0)^{3}}-\frac{2V^{'}(\phi_0)^{3}\delta}{V(\phi_0)^{4}}+\frac{V^{'}(\phi_0)^{2}V^{'''}(\phi_0)}{V(\phi_0)^{3}}
-\frac{2V^{''}(\phi_0)V^{'''}(\phi_0)}{3V(\phi_0)^{2}}-\frac{V^{''''}(\phi_0) V^{'}(\phi_0)}{2V(\phi_0)^{2}},\\
\displaystyle \cdots \cdots \cdots \cdots \cdots \cdots\cdots \cdots \cdots \cdots \cdots \cdots\cdots \cdots \cdots
   \end{array}\ee
Here $V(\phi_0),V^{'}(\phi_0),V^{'''}(\phi_0),V^{''''}(\phi_0)\neq 0,V^{''}(\phi_0)=0$ and $V(\phi_0),,V^{'''}(\phi_0),V^{''''}(\phi_0)\neq 0,V^{'}(\phi_0),V^{''}(\phi_0)=0$ 
are two limiting situations which signifies the {\it inflection point} and {\it saddle point} inflationary setup. For details, see Ref.~\cite{Choudhury:2013iaa}.


\end{document}